# Automated Molecular Beam Epitaxy Growth of AlAs/GaAs Bragg mirrors

Pierre Gadras[*], Léo Bourdon, Antoine Fées, Karim Ben Saddik, Guilhem Almuneau and Alexandre Arnoult

LAAS-CNRS, Université de Toulouse, CNRS, INSA, 31400 Toulouse, France
pgadras@laas.fr; almuneau@laas.fr

**Abstract**

In-situ measurement is a key feature to better understand and precisely control the growth of complex structures, such as VCSEL. In this work, we are showing the automated growth by molecular beam epitaxy of a GaAs/AlAs Distributed Bragg Reflector (DBR) centered at 940 nm, feedback controlled with in-situ spectral reflectance measurement and realized without preliminary accurate cell flux calibrations. This method relies on the measurements of the optical indices' dispersion over wide spectral range (400 – 1400 nm) at growth temperature (600°C) of $Al_xGa_{1-x}As$ alloys. To do so, in-situ reflectance measurement is used combined with ex-situ layer thickness and composition measurement by X-ray diffraction. The DBR sample is characterized by X-ray diffraction and Fourier transform infrared reflectance, resulting in a deviation of 0.2 nm of the stop-band central-wavelength compared to the targeted one.

## 1. Introduction

In Molecular Beam Epitaxy (MBE), calibration is traditionally performed through a cycle of growth and ex-situ characterizations, such as X-ray diffraction (XRD). While in-situ tools like ion gauge flux measurements have limited accuracy [1], achieving precise growth of complex multilayer device structures still requires at least one calibration sample to determine the exact growth rate for each layer. Ideally, growth rate measurements could be performed during the epitaxy process, allowing real-time adjustments to ensure accuracy.

As a result, in-situ measurement techniques have become increasingly appealing in semiconductor fabrication. In particular, optical reflectance and transmittance measurements of samples during growth are widely used to control deposition rates in techniques such as Metal-Organic Chemical Vapor Deposition (MOCVD) [2] and Physical Vapor Deposition (PVD) [3].

In Molecular Beam Epitaxy (MBE), various approaches have been proposed to monitor optical component growth [4–9]. However, to our knowledge, no commercially available tool effectively addresses this issue. Consequently, the data in the literature originates from diverse measurement techniques and setups. This results in scattered performance reports that do not provide precise knowledge of the refractive indices over full compositional and spectral ranges, as is the case with AlGaAs. Moreover, modern MBE systems have advanced significantly and are now predominantly controlled by centralized and standardized systems, such as CrystalXE software from Riber. This development enables more detailed in-situ monitoring techniques and integrated control strategies.

In this work, for achieving most accurate in-situ assessment of layer thicknesses, we measured under MBE growth conditions the refractive indices (n, k) of the $Al_xGa_{1-x}As$ alloys across the whole composition range and over a wide spectral VIS-NIR range based on reflectometry in-situ measurements. Particular attention is given to the determination of the confidence interval. Based on these preliminary calibrations, we are demonstrating the automated controlled growth of a GaAs/AlAs Distributed Bragg Reflector (DBR) centered at 940 nm, using a feedback-controlled process from in-situ spectral reflectance measurement. Remarkably, this was achieved without any prior accurate cell flow calibrations.

## 2. Theoretical background

This work is based on the method proposed by Breiland and Killeen [4], which provides the theoretical framework. According to their study, the evolution of reflectance over time during the growth of a semiconductor thin film can be described by a damped cosine equation (see equation (8) in [4]). Using this formulation, the optical indices $(n, k)$ of the growing thin film can be derived from the reflectance variations as follows:

$$n = \frac{\lambda}{2 * G * T * cos\theta_1} \quad (1)$$

$$k = \frac{\lambda * \gamma}{4\pi * G * cos\theta_1} \quad (2)$$

With $\lambda$ the measured wavelength, $G$ the growth rate of the thin film (assumed constant) and $\theta_1$ the incident angle in the growing layer. The expressions of (1) and (2) as a function of the absolute incident angle ($\theta_0$) in vacuum (see Fig. 2) are given in the supplementary document. The period of the oscillation ($T$) and damping coefficient ($\gamma$) are deduced from the measured reflectance, as illustrated in Fig. 1 (measurement details are given in section 3.b).

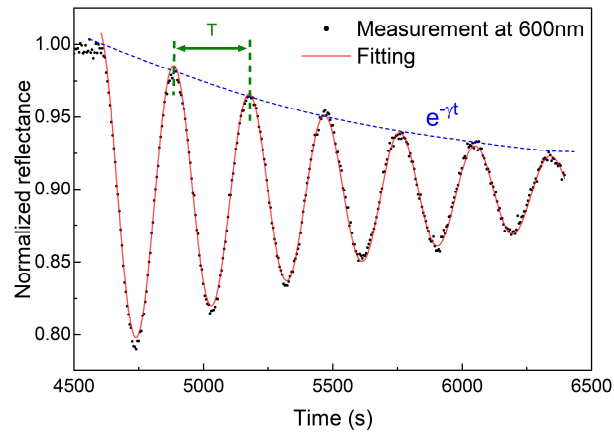

Fig. 1 : In-situ reflectivity during the growth of 0.5 μm of $Al_{0.53}Ga_{0.47}As$ measured at 600 nm and 600°C with an incident angle $\theta_0$ of 20°. Equation (1) fitted function (red solid line) allows the extraction of the oscillation period $T$ and the damping coefficient $\gamma$ (blue dashed line).

In order to ascertain the values of n and k under growth conditions, the reflectance time variation $R(t)$ was measured during the growth of a homogeneous AlGaAs thin film. Subsequently, the layer thickness and its corresponding growth rate, $G$, were measured ex-situ on the same sample at room temperature using X-Ray

Reflectometry (XRR). The reflectance measurement is fitted to extract the parameters $T$ and $\gamma$, thereby enabling the independent determination of the optical indices n and k for each wavelength, without requiring the assumption of an index model.

In the few data in the literature on AlGaAs, optical index measurements at high temperatures have been obtained using different methodologies, but they show discrepancies (see S.2). In this study we focused on quantifying the global uncertainty of our measurement, and identifying the impacts in terms of errors on the determination of the optical index. To assess the reliability of the deduced optical indices, we express the standard deviation $\sigma$ with the following equations by deriving expressions ( 1 ) and ( 2 ) with the following general formula [10] :

$$\sigma(f) = \sqrt{\sum_i \left(\frac{\partial f}{\partial x_i} * \sigma(x_i)\right)^2} \qquad (3)$$

With $\sigma(x_i)$ representing the standard deviation for each parameter $x_i$

We get:

$$\sigma(n) = n * \sqrt{\left(\frac{1}{T}*\sigma(T)\right)^2 + \left(\frac{1}{\lambda}*\sigma(\lambda)\right)^2 + \left(\frac{1}{G}*\sigma(G)\right)^2 + (tan\theta_1 * \sigma(\theta_1))^2} \qquad (4)$$

$$\sigma(k) = k * \sqrt{\left(\frac{1}{\gamma}*\sigma(\gamma)\right)^2 + \left(\frac{1}{\lambda}*\sigma(\lambda)\right)^2 + \left(\frac{1}{G}*\sigma(G)\right)^2 + (tan\theta_1 * \sigma(\theta_1))^2} \qquad (5)$$

We take $\sigma(\theta_0) = 0.5°$ based on the MBE reactor geometry, from which we derive $\sigma(\theta_1)$ as described in the supplementary document. Since the two spectrometers used in this work had different resolutions, we use the standard deviation accordingly taken from the manufacturer's datasheet: $\sigma(\lambda) = 0.03nm$ for UV-VIS spectrometer, and $\sigma(\lambda) = 0.6nm$ for the NIR one; the transition wavelength being around 1000nm. The remaining standard deviations, $\sigma(T)$, $\sigma(\gamma)$, and $\sigma(G)$, are experiment-dependents and are determined from the covariance matrix derived from the fitting procedure of the in-situ reflectance oscillations and the ex-situ growth rate measurement, respectively.

When the influence of each parameter on the overall standard deviation is compared, it becomes clear that errors in growth rate estimation are the predominant source of uncertainty in determining the refractive index $n$. Conversely, the determination of the $\gamma$ parameter significantly contributes to the error in the optical coefficient $k$. This is due to the difficulty of numerical resolution to fit the decreasing envelope of the signal, especially in the case of low absorption. Detailed error contributions for each parameter are listed in the Table S1.

Moreover, we estimated a potential error on the value of indices $n$ in the order of $\Delta n \sim 0.0004$ (equation (IV.3) in [5]) due to the uncertainty of the temperature measurement $\Delta T = \pm 1°C$, which it is considered negligible.

3. **Materials and methods**
    a. MBE growth and sample structure

The growths were done on standard 2-inch semi-insulating GaAs (001) wafers using a RIBER MBE 412 reactor equipped with Knudsen effusion cells for group-III elements and a cracker cell for arsenic. The growth temperature for all samples was set to 600±1°C, monitored via absorption band-edge thermometry (BandiT) on the bare wafer prior to growth, and stabilized during the process using thermocouple-based PID control loop.

The V/III beam equivalent pressure ratio was set at 3.2 for $Al_xGa_{1-x}As$ and at 2 for $GaAs$. In-situ spectral reflectance measurements were performed using a homemade setup, as illustrated in Fig. 2. This setup, mounted on the 20°-off axis symmetric viewports of the MBE reactor, consists of a halogen white light source, a linear polarizer, a beam splitter and a lens system for collecting and coupling light into two fibers towards the two spectrometers: ULS2048CL-EVO for UV-VIS (400–1100 nm) and N256-17-EVO for NIR (1000–1700 nm) from Avantes. The spectra from both spectrometers are acquired in synchronization with the substrate rotation and are merged and processed using a LabVIEW code linked to the MBE main control software (CrystalXE).

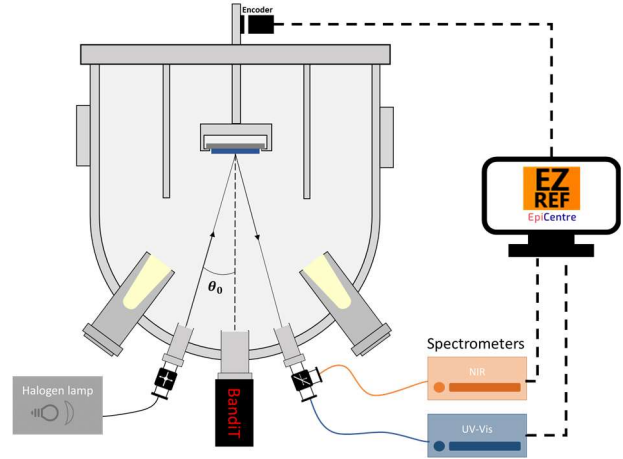

Fig. 2 : Schematic of the In-situ spectral reflectance measurement setup on MBE reactor

The sample stack consists of two sections: the first set of layers enables in-situ measurement of refractive indices, while the second allows precise ex-situ growth rate measurements. In the first section of the stack, to deduce the optical indices at high temperatures, we measure in-situ reflectance during the growth of a 500 to 1000 nm-thick layer of $Al_xGa_{1-x}As$ depending on the composition. This relative thick layer allows multiple interference oscillations, resulting in a more accurate estimation of the period, particularly for longest wavelengths, while minimizing the effect of the transient flow burst at the initial stage of the layer growth just after opening the cell shutter. For measurements carried out on GaAs films, we include a thin optical index stepping layer of AlAs (30 nm), in order to clearly observe interference reflectance oscillations. The second section of the stack consists of two layers (10 nm and 150 nm) with different aluminum compositions, for which we measure their thicknesses using XRR, allowing us to accurately derive the growth rates on the same growth run.

b. *In-situ* characterization

*In-situ* normalized reflectance measurements were carried out with respect to the initial deoxidized substrate surface spectrum ($R_0$), typically obtained from the GaAs substrate at the growth temperature. During the growth process, clear patterns of constructive and destructive interference were observed, as shown Fig. 3. These interference fringes serve as a critical tool for monitoring layer thickness and growth dynamics in real time. At 700 nm, an abrupt change in reflectivity is observed due to the band gap. To obtain a wide spectral range measurement, data from the two spectrometers were merged numerically. In the image, a darker line appears at 1010 nm, corresponding to the merging wavelength, where both spectrometers exhibit reduced sensitivity. Additionally, in the infrared region beyond 1200 nm, a significant reduction in the signal-to-noise ratio (SNR) was observed. This degradation is primarily due to the blackbody emission from the heated furnace behind the substrate, which introduces background noise and reduces the quality of the reflectance signal.

Despite this, the reflectance signals were successfully fitted over the wavelength range of 400–1400 nm. The non-linear least squares method was employed to fit each wavelength, like shown in Fig. 1, using the damped cosine expression as follow:

$$\frac{R(t)}{R_0} \approx A - B * cos\left(\frac{2\pi}{T}t + \phi\right) * e^{-\gamma t} \quad (6)$$

With fitted parameters $A, B, T, \phi, \gamma$, but as previously mentioned only the parameters $T$ and $\gamma$ are used in equation ( 1 ) and ( 2 ) to deduce the optical coefficient of the material with growth rate previously measured.

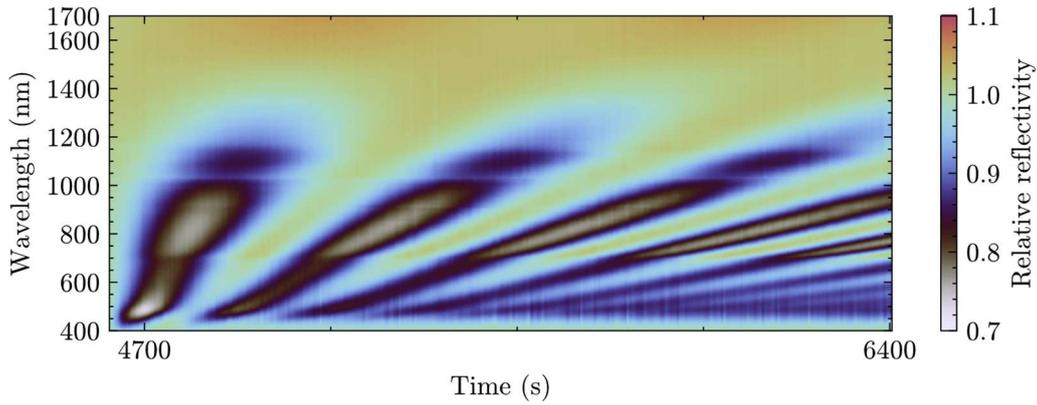

Fig. 3 : In-situ spectral reflectance measurement during the growth of $Al_{0.53}Ga_{0.47}As$ 500-nm thick layer on $GaAs$

c. X-ray characterization

X-ray measurements were performed using a Bruker D8 system. Growth rate values were determined from XRR measurements, accounting for temperature-induced dilation differences [11]. While aluminum concentrations were derived from XRD measurements around the (004) reflection. These parameters were simultaneously fitted using the Diffract Leptos software. The corresponding uncertainties were also derived from the software; however, they account only for fitting uncertainty and not measurement uncertainty. Moreover, we estimated on-wafer thickness inhomogeneity (0.5%) which was lower than XRR fitting uncertainty (1%), meaning that thickness non-uniformity across the wafer is not a significant factor.

As demonstrated in [12], growth rate measurements using XRR tend to be significantly more precise than those obtained through XRD, provided an appropriate stack is used. Precision can be further enhanced by increasing the thickness of the measured layer, as this results in more Pendelössung fringes (interference between the incident wave and the interfaces) within the measurement angular range, offering a larger dataset for fitting. However, a thicker stack reduces X-ray interaction with the underlying layer, decreasing fringes amplitudes and measurement contrast. A good compromise was found with thickness of 150 nm, as shown in Fig. 4, where $Al_{0.53}Ga_{0.47}As$ layer exhibits more than 40 oscillations in the measurement range, enabling growth rate determination with errors as low as 0.12%.

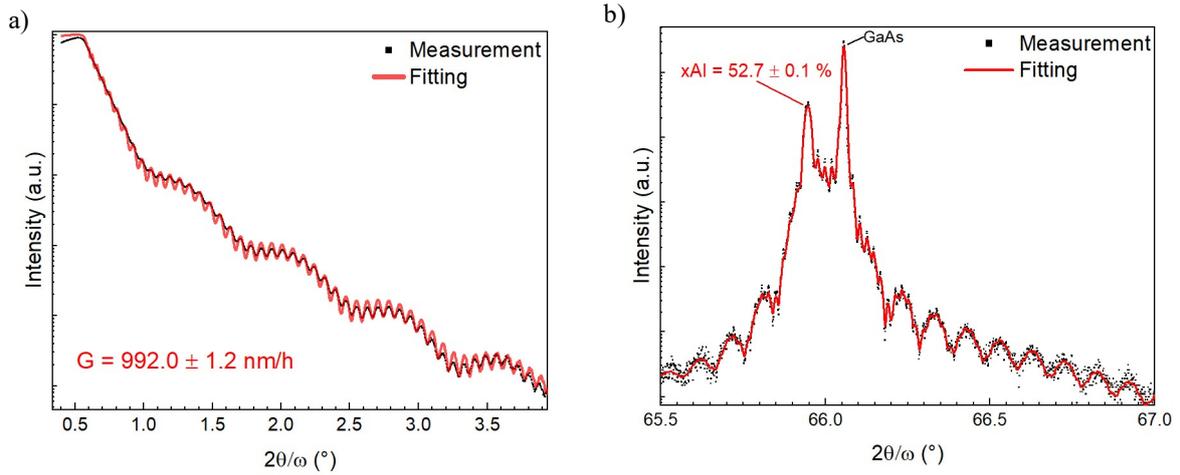

Fig. 4 : X-ray measurement *a)* in reflection and *b)* at (004) diffraction angle for $Al_{0.53}Ga_{0.47}As$ sample

## 4. Optical indices determination

In Fig. 5, we present the refractive index and extinction coefficient dispersion over the range of 400-1400 nm, at $600 \pm 1°C$ for $Al_xGa_{1-x}As$ compositions of 0, 0.31, 0.53, 0.92, and 1. Compared to room-temperature data, an increase in temperature results in a redshift and a broadening of the spectral features. Unlike most of the limited data available in the literature, which were obtained using spectroscopic ellipsometry, our measurements do not require a model or consideration of an oxide layer, thereby reducing potential sources of errors. Confidence bounds are shown by shaded areas around the curves, based on the uncertainty assessment detailed in equation ( 4 ) and ( 5 ). Errors in the refractive index (n) are in the order of 0.05, while errors in the extinction coefficient (k) are around 0.001. These low-error measurements make the data highly valuable for growth control-loop feedback, as detailed in the next section. To our knowledge, this is the first study to report systematic measurements of the optical constants of multiple $Al_xGa_{1-x}As$ compositions at growth temperature over a broad wavelength range.

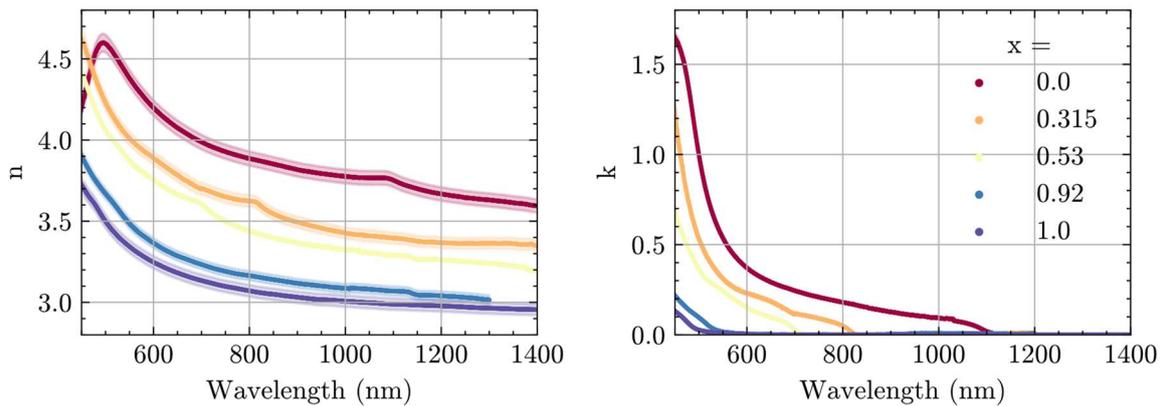

Fig. 5 : *In-situ* measured refractive indices ($n$) and extinction coefficients ($k$) of $Al_xGa_{1-x}As$ at 600°C between 450 nm and 1400 nm. Shaded areas correspond to the estimated vertical errors (95%).

Divergences in the fit procedure are observed only at long wavelengths, as it can be seen for the $Al_{0.92}Ga_{0.08}As$ case for which the reflectivity oscillations cannot be well fitted around 1300 nm. As a result, the data above this wavelength have been excluded for this sample. A possible explanation is the effect of ambient heating and blackbody emission, which reduces the signal-to-noise ratio (SNR) of the reflectance measurements, as shown in the upper part of Fig. 3. Such divergence has also been reported in the literature for *in-situ* optical index measurements [13,14]. In addition, due to the increase of the oscillation period of the reflectance signal at higher wavelengths, fewer oscillation periods are available for fitting the same thickness, which reduces the precision of the measurement.

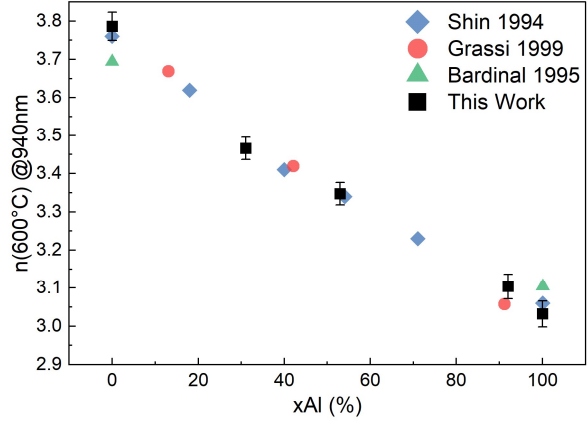

Fig. 6 : Measured optical index variation of $Al_xGa_{1-x}As$ at 940 nm under 600°C as a function of Al concentration, compared with values reported in the literature from [5,15,16].

In the spectral range relevant to the targeted DBR designed at 940 nm, which corresponds to a transparent region for $Al_xGa_{1-x}As$ with x>0.10, we observed good agreement with the limited available literature data on narrow wavelength range as shown in Fig. 6. But as shown in Fig. S1 the literature data shows significant discrepancies in optical index measurements in the visible spectral range. These variations can be attributed to differences in doping levels, consideration of a native oxide, discrepancies in the temperature determination, or even measurement techniques.

## 5. Automated feedback-controlled DBR growth

### a. Methodology

To implement our automated growth strategy, we firstly simulated the in-situ spectral reflectance during the growth of an ideal DBR using the scattering matrix method (SMM) [17] with previously measured optical index wavelength dispersions. Next, at the end of each quarter-wavelength layer, we calculated the wavelength for which the reflectance reaches a maximum (for the high index layer), or a minimum (for the low index layer). This wavelength is referred as the optimal wavelength [5,18]. Resulting wavelengths are plotted in Fig. 7 superimposed to the reflectivity map vs time and wavelength. The first derivative of reflectance with respect to time was used to detect these extrema.

### b. Experiment

For the growth of the DBR, the gallium and aluminum effusion cells were regulated to temperature setpoints corresponding to a growth rate of approximately 0.5 µm/h, based on reactor history. However, these values were not previously calibrated using ion flux gauge.

To control the shutter automated switching, we implemented a script directly in the MBE reactor control software (CrystalXE). This script reads the relative reflectance measurements at the corresponding optimum wavelengths over the entire structure, records the last values and toggles the reactor cell shutter between Ga and Al at the moment a maximum is detected, or respectively a minimum, in the time variation of the reflectance. Extrema were detected with a method inspired from the 5-points method discussed in [19], which involves comparing the last five values recorded. An extremum is identified when the last value is either lower (minimum) or higher (maximum) than the previous four. Fig. 7 shows the measured in-situ reflectance map of the sample. Wavelength range was reduced for clarity. As expected for a DBR, the reflectivity increases around the optimal wavelength and the stop band narrows when more layers are stacked.

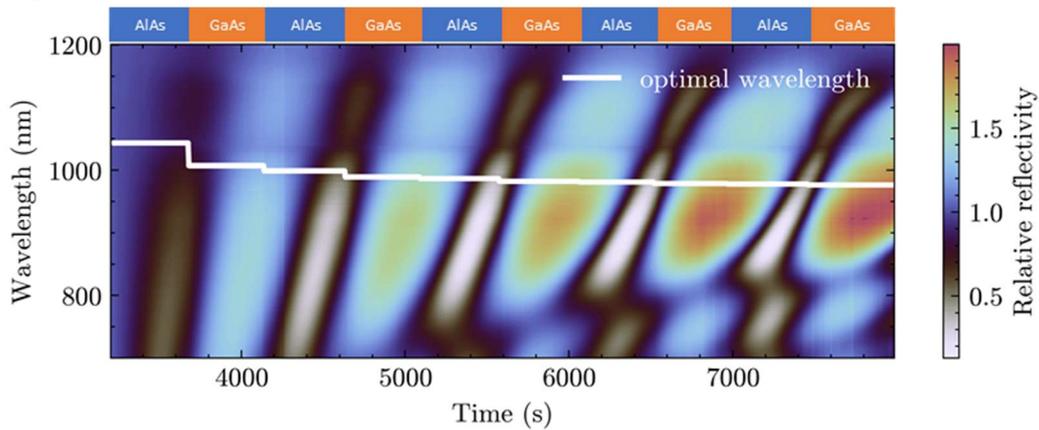

Fig. 7 : In-situ reflectance measurement during the growth of a feedback-controlled 5-period DBR. The superimposed white line indicates the calculated optimal wavelength for the DBR structure.

c. Results and discussion

HRXRD measurement around (004) reflection was performed on this DBR sample, as shown in Fig. 8a. The measured diffraction signal (solid black line) deviates significantly from the simulated diffractogram for ideal $\lambda/4$-layer thicknesses (dashed blue line). By fitting the thickness of each of the 10 layers independently, we obtained a nearly perfect agreement with the experimental diffractogram, shown by the dashed red line. Fig. 8b illustrates the statistical dispersion of the fitted layer thicknesses for the two materials. The averaged GaAs and AlAs layer thicknesses deviate from the theoretical values by more than 7% for both materials.

Fourier Transform Infrared (FTIR) measurement was conducted utilizing a Bruker Vertex 70 equipped with a dual reflection configuration. The setup angle of incidence is 12° and is taken into account in the simulations. Reflectivity measurements in Fig. 9, agree relatively well with the simulation for ideal $\lambda/4$ layers, except for the short wavelength sidelobes. However, the reflectivity calculated with the thicknesses deduced from XRD perfectly matched over the whole measurement range. Notably, within the stop-band — defined by the range between the 90% reflectance points — the center wavelength aligns closely with the theoretical value. These results highlight the phase compensation phenomenon, a well-documented effect associated with this control method [20].

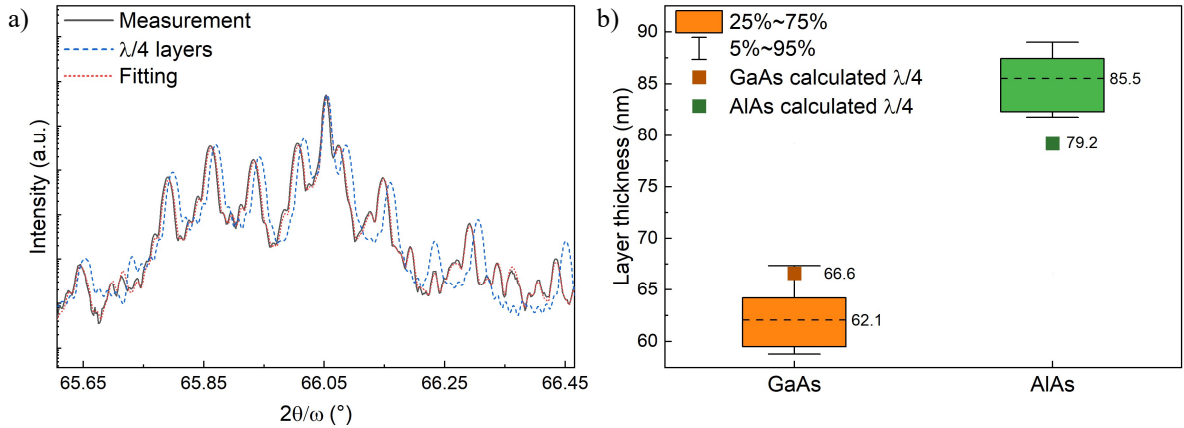

Fig. 8 : a) HRXRD measurement of the GaAs/AlAs DBR: The solid black line represents the measurement, the dashed blue line corresponds to the simulation with ideal λ/4 layers, and the dotted red line indicates the fitted data. b) Statistical distribution of layer thicknesses compared to the ideal λ/4 thickness.

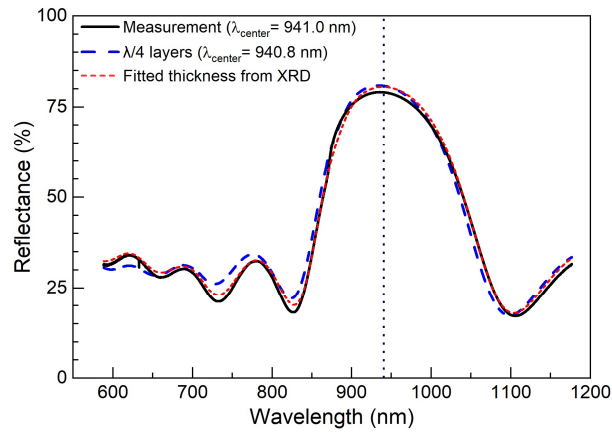

Fig. 9 : Comparison between reflectance measurement of automated DBR growth (Full black line)) with SMM simulation with XRD thicknesses (Short dashed red line) and quarter-wavelength thicknesses (Long dashed blue line).

## 6. Conclusion

In this study, we implemented systematic in-situ spectral measurements to determine the optical indices of $Al_xGa_{1-x}As$ materials for various compositions at the growth temperature of 600°C and their wavelength dispersion between 450 nm and 1400 nm. This paper's indices data complements the existing literature. It also shows good agreement with the limited data available in the literature within the transparent spectral range, nevertheless discrepancies are evident across all referenced studies in the short visible region. These measurements were then used to calibrate and feedback-loop control the growth of a GaAs/AlAs distributed Bragg reflector (DBR) centered at 940 nm. Excellent spectral matching to the target wavelength was achieved with the automatically grown 5-periods DBR. This despite the fact that XRD measurements revealed significant deviations in layer thicknesses compared to ideal quaterwave values. This discrepancy is attributed to self-phase compensation effects introduced by the feedback control loop.

Based on the accurate knowledge of the optical indices at growth temperature, the proposed method has the potential to significantly reduce the number of calibrations runs required to achieve the targeted photonic devices.

Future work will focus on applying this method to the growth of complex III-V epitaxial structures, such as VCSELs, paving the way for further advancements in the full automation of MBE fabrication processes.

## SUPPLEMENTARY MATERIAL

See the supplementary material can be found online at …

## ACKNOWLEDGMENTS

This research was supported by the EPICENTRE, Joint laboratory between CNRS-LAAS and RIBER, the European Union PhotoGeNIC project under the grant agreement no. 101069490, and the Région Occitanie through the Défi Clé PV-STAR.  This work was realized in the LAAS-CNRS micro and nanotechnologies platform, a member of the French RENATECH network.

## AUTHOR DECLARATIONS

Conflict of Interest

The authors have no conflicts to disclose.

Author Contributions

~~Arslan Ahmed: Conceptualization (equal); Investigation (equal); Methodology (equal); Visualization (equal); Writing – original draft (equal). Nurul Syafeeqa Ishak: Investigation (equal); Software (equal); Validation (equal). Fauziahanim Che Seman: Conceptualization (equal); Funding acquisition (equal); Methodology (equal); Project administration (equal); Supervision (equal); Validation (equal); Writing – review & editing (equal). See Khee Yee:: Funding acquisition (lead); Project administration (equal); Supervision (equal). Sajjad Ahmed: Resources (equal).~~

## DATA AVAILABILITY

The data that support the findings of this study are available from the corresponding author upon reasonable request.

Supplementary material for
Automated Molecular Beam Epitaxy Growth of AlAs/GaAs Bragg mirrors

Pierre Gadras*, Léo Bourdon, Antoine Fées, Karim Ben Saddik, Guilhem Almuneau and Alexandre Arnoult

LAAS-CNRS, Université de Toulouse, CNRS, INSA, 31400 Toulouse, France

*pgadras@laas.fr; almuneau@laas.fr

1. **Angle of incidence in the thin film**

As shown in equations ( 1 ) and ( 2 ), optical indices coefficients are linked to the incident angle in the growing thin film medium ($\theta_1$). But as we only know the absolute angle of incidence ($\theta_0$), we need to link those two. We develop this link in the following equations:

$$n_1 * sin\theta_1 = n_0 * sin\theta_0$$

With incident medium $n_0 = 1$ :

$$cos\theta_1 = \sqrt{1 - \frac{sin\theta_0^2}{n_1^2}} \qquad (7)$$

By applying **Erreur ! Source du renvoi introuvable.** with ( 1 ) and ( 2 ), we can express:

$$n = \frac{\lambda}{2 * G * T * cos\theta_1} = \sqrt{(sin\theta_0)^2 + \left(\frac{\lambda}{2 * G * T}\right)^2} \qquad (8)$$

$$k = \frac{\lambda * \gamma}{4\pi * G * cos\theta_1} = \frac{\lambda * \gamma}{4\pi * G * \sqrt{1 - \frac{sin^2\theta_0}{n_1^2}}} \qquad (9)$$

And so, we observe that in the measured *k* value depend effectively of the measured *n* value. However, as in this case $\theta_0 = 20°$, the value of $\left(\frac{sin\theta_0}{n_1}\right)^2$ is around 0.01 and so it as a weak impact.

Now we can express the corresponding uncertainty estimation for expression ( 4 )( 5 ), respectively of the absolute incident angle ($\theta_0$) :

$$\sigma(n) = \frac{1}{n} * \sqrt{\left(\frac{\lambda^2}{4G^2T^3} * \sigma(T)\right)^2 + \left(\frac{\lambda}{4G^2T^2} * \sigma(\lambda)\right)^2 + \left(\frac{\lambda^2}{4G^3T^2} * \sigma(G)\right)^2 + \left(\frac{sin2\theta_0}{2} * \sigma(\theta_0)\right)^2} \qquad (10)$$

Moreover, we can express $\theta_1$ as:

$$\theta_1 = sin^{-1}\left(\frac{sin\theta_0}{n_1}\right) \qquad (11)$$

And so, we can develop the uncertainty estimation $\sigma(\theta_1)$ from $\sigma(\theta_0)$ and, previously determined $\sigma(n)$ and use this formulation to easily calculate $\sigma(k)$:

$$\sigma(\theta_1) = \frac{1}{n\sqrt{1-sin^2\theta_1}} * \sqrt{\left(\frac{sin\theta_0}{n} * \sigma(n)\right)^2 + (cos\theta_0 * \sigma(\theta_0))^2} \qquad (12)$$

2. **Optical indices comparison with literature**

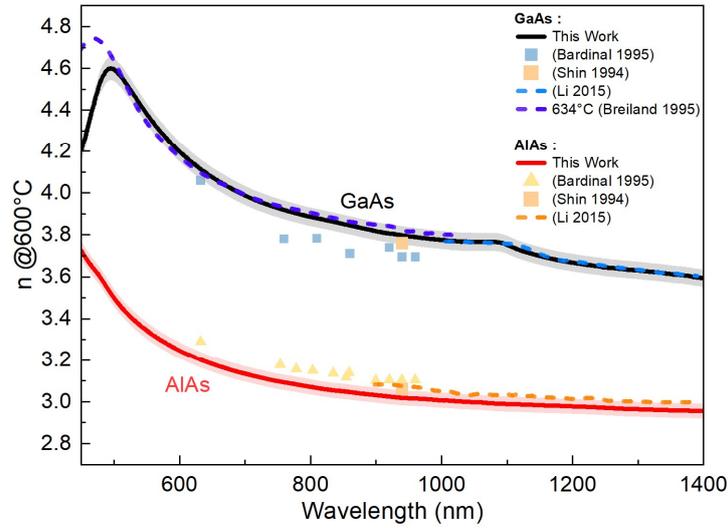

Fig. S1: Comparison of refractive index measurement of AlAs and GaAs at 600°C between this work and literature from [4,16,15,8]

3. **Uncertainty contributions**

Table S1 presents the calculated uncertainties at the wavelength of 940 nm and the relative contribution of each parameter to the final computed values. The k value for AlAs is not measurable at this wavelength; thus, the corresponding standard deviation is omitted in this table.

$$x_{i\,relative contribution} = \frac{\left(\frac{\partial f}{\partial x_i} * \sigma(x_i)\right)^2}{\sigma(f)^2}$$

Table. S1 : Relative contribution of standard deviation from different measurement variables to the determination of the refractive index (n) and extinction coefficient (k) for GaAs and AlAs at 940 nm at a growth temperature of 600°C.

|  | GaAs | | AlAs |
|---|---|---|---|
|  | n | k | n |
| $\sigma(f)/f$ | 0.741% | **1.095%** | **0.610%** |
| $G$ contribution | 99.338 % | 46.202 % | 98.525 % |
| $T \vee \gamma$ contribution | 0.077% | 53.762 % | 0.028 % |
| $\theta_1$ contribution | 0.071 % | 0.036 % | 0.257 % |
| $\lambda$ contribution | 0.002 % | 0.001% | 0.002 % |
| T° contribution for $600 \pm 1°C$ | 0.513 % | n.a. | 1.187 % |